\documentclass{article}
\usepackage{graphicx}
\usepackage{comment}
\usepackage{amssymb} 
\usepackage{amsmath}
\newtheorem{theorem}{Theorem}
\newtheorem{lemma}{Lemma}
\newtheorem{proposition}{Proposition}
\newtheorem{corollary}{Corollary}
\newtheorem{example}{Example}
\date{}

\title{On OBDDs for CNFs of bounded treewidth}
\author{Igor Razgon\\ Department of Computer Science and Information Systems,\\ Birkbeck, University of London\\
        igor@dcs.bbk.ac.uk}
\begin{document}
\maketitle
\begin{abstract}
Knowledge compilation is a rewriting approach to propositional knowledge representation.
The `knowledge base' is initially represented as a {\sc cnf} for which many important types of
queries are {\sc np}-hard to answer. Therefore, the {\sc cnf} is compiled into another
representation for which the minimal requirement is that the clausal entailment query
(can the given partial assignment be extended to a complete satisfying assignment?)
can be answered in a polynomial time \cite{DerMar}. Such transformation can result in exponential
blow up of the representation size. A possible way to circumvent this issue is to
identify a structural parameter of the input {\sc cnf} such that the resulting transformation
is exponential in this parameter and polynomial in the number of variables.
A notable result in this direction is an $O(2^kn)$ upper bound on the size of
Decomposable Negation Normal Form ({\sc dnnf}) \cite{DarwicheJACM}, where $n$ is the
number of variables of the given CNF and $k$ is the treewidth of its primal graph. Quite recently
this upper bound has been shown to hold for Sentential Decision Diagrams ({\sc sdd}) \cite{SDD}, 
a subclass of {\sc dnnf} that can be considered as a generalization of the famous Ordered Binary Decision Diagrams
({\sc obdd}) and shares with the {\sc obdd} the key nice features (e.g. poly-time equivalence testing). Under the 
treewidth parameterization, the best known upper bound for an {\sc obdd} is $O(n^k)$ \cite{VardiTWD}. 
A natural question is whether, similarly to {\sc sdd}, a fixed parameter upper bound holds for {\sc obdd}.

We provide a negative answer to the above question. In particular, for every fixed $k$, 
we demonstrate an infinite class of {\sc cnf}s of the primal graph treewidth at most $k$ for which the 
{\sc obdd} size is $\Omega(n^{k/4})$, essentially matching the upper bound of \cite{VardiTWD}.
This result establishes a \emph{parameterized} separation of {\sc obdd} from {\sc sdd}. We further show that
the considered class of instances can be transformed into one for which the {\sc obdd} size is at least $n^{\Omega(\log n)}$
and the {\sc sdd} size is $O(n^3)$ thus separating {\sc obdd} from {\sc sdd} in the \emph{classical sense}.

We also provide a more optimistic version of the $O(n^k)$ upper bound for the {\sc obdd} showing
that it in fact holds when $k$ is the treewidth of the incidence graph of the given {\sc cnf}.
\end{abstract}

\section{Introduction}
Knowledge compilation is a rewriting approach to propositional knowledge representation.
The `knowledge base' is initially represented as a {\sc cnf} or even as a Boolean circuit. 
For these representations many important types of queries are {\sc np}-hard to answer.
Therefore, the initial representation is compiled into another one 
for which the minimal requirement is that the clausal entailment query
(can the given partial assignment be extended to a complete satisfying assignment?)
can be answered in a polynomial time \cite{DerMar}. Such transformation can result in exponential
blow up of the representation size. A possible way to circumvent this issue is to
identify a structural parameter of the input {\sc cnf} such that the resulting transformation
is exponential in this parameter and polynomial in the number of variables.
A notable result in this direction is an $O(2^kn)$ upper bound on the size of
Decomposable Negation Normal Form ({\sc dnnf}) \cite{DarwicheJACM}, where $n$ is the
number of variables of the given {\sc cnf} and $k$ is the treewidth of its primal graph.
Quite recently, the same upper bound has been shown to hold for Sentential Decision Diagrams ({\sc sdd}) \cite{SDD},
a subclass of {\sc dnnf}  
that can be seen as a generalization
of the famous Ordered Binary Decision Diagrams ({\sc obdd}) and 
shares with the {\sc obdd} the key nice features (e.g. poly-time equivalence testing). It is known that a {\sc cnf} of treewidth
$k$ can be compiled into an {\sc obdd} of size $O(n^{k})$ \cite{VardiTWD}. A natural question is whether 
{\sc obdd}, similarly to {\sc sdd}, admits a fixed-parameter upper bound of form $f(k)n^c$ for some constant $c$. 

In this paper we provide a negative answer to this question. In particular, we demonstrate an infinite class of
{\sc cnf}s of the primal graph treewidth at most $k$ for which the {\sc obdd} size is at least
$f(k)n^{k/4}$ where $f$ is a function exponentially small in $k$. In other words, we show
that the {\sc obdd} size of these {\sc cnf}s is $\Omega(n^{k/4})$ for every fixed $k$.
This result provides a \emph{parameterized} separation from {\sc sdd} and essentially matches the upper bound of \cite{VardiTWD}. 
In fact, this result shows impossibility of not only a fixed-parameter upper bound, but also 
of a sublinear dependence on $k$ in the base of the exponent or even of 
an exponent $k/C$ for some large constant $C$. Moreover, a corollary of this result is that there is
an infinite class of instances (obtained, roughly speaking, by setting $k=\log n$) on which the {\sc obdd} 
size is at least $n^{\Omega(\log n)}$, while the {\sc sdd} size is $O(n^3)$ thus separating {\sc obdd} 
from {\sc sdd} in the \emph{classical sense}.

Our second result is `strengthening' of the upper bound $O(n^{k})$ of \cite{VardiTWD} by showing that it
holds if $k$ is the treewidth of the \emph{incidence} graph of the given {\sc cnf} thus extending the 
upper bound to the case of sparse {\sc cnf}s with large clauses.

In order to obtain the parameterized lower bound, we introduce a notion of \emph{matching width} of a graph
and prove that if a {\sc cnf} $F$ of the considered class
has matching width $r$ of the primal graph then for any ordering of the variables of $F$
there is a prefix $S$ such that the number of distinct functions that can be obtained 
from $F$ by assigning the variables of $S$ is at least $2^r$.
This will immediately imply that any {\sc obdd} realizing
$F$ will have at least $2^r$ nodes. Finally we will prove that the matching width of the considered
{\sc cnf}s is $\Omega (logn*k)$. Substituting this lower bound instead $r$ will get the
desired lower bound for the {\sc obdd} size. 

Similarly to the case of primal graph, the upper bound is obtained by showing that if \emph{pathwidth}
of the incidence graph of the given {\sc cnf} is at most $p$ then this {\sc cnf} can be compiled into 
an {\sc obdd} of size $O(2^pn)$. Then the $O(n^k)$ upper bound is obtained using a well known relation
$p=O(k*logn)$ between the treewidth and the pathwidth of the given graph. The approach to obtain the 
$O(2^pn)$ bound is similar to \cite{VardiTWD}: variables are ordered 'along' the path decomposition
and it is observed that the for each prefix the number of functions caused by assigning the 'previous'
variables is $O(2^p)$. The technical difference is that in our case the bags of the path decomposition
include clauses and this circumstance must be taken into account. 

The proposed results contribute to a large body of existing results concerning the space complexity
of {\sc obdd}s. To begin with, there are many results concerning the complexity of {\sc obdd}s for 
\emph{particular} classes of Boolean functions, see e.g. the book \cite{WegBook} and the survey \cite{WegSurvey}. 
The space complexity of {\sc obdd} remains polynomial if parameterized by the treewidth of a \emph{circuit} representing the given 
function \cite{OBDDTWJha}, however the dependence on the treewidth becomes double exponential. 
A fixed-parameter upper bound can be achieved if tree of {\sc obdd}s is used instead of a single
{\sc obdd} \cite{McMillan94,SubbaTree}. In the complexity theory the {\sc obdd} is classified
as the \emph{oblivious read-once branching program}, see the book \cite{Yukna} for the results concerning the
complexity of branching programs on particular classes of formulas

The proposed lower bound also contributes to the understanding of relationship between {\sc obdd}
and {\sc sdd}. Other results in this direction are \cite{SDDvsOBDD} showing an exponential separation
between {\sc sdd} and {\sc obdd} based on the same order of variables (the order of variables for
{\sc sdd} is defined as the order of visiting the corresponding nodes of the underlying \emph{vtree}
by a left-right tree traversal algorithm) and \cite{DynSDD} empirically showing that conceptually similar 
heuristics produce {\sc sdd}s orders of magnitude smaller than {\sc obdd}s.

The rest of the paper is structured as follows.
The next section introduces the necessary background. 
The section after that proves the lower bound, the proofs of auxiliary statements 
are provided in the two following sections. 
Then follows the section presenting the
upper bound for the parameterization
by the treewidth of the incidence graph. 

\section{Preliminaries} \label{prelim}
The structure of this section is the following. First, we introduce
notational conventions. Then we define the {\sc obdd} and specify the approach
we use to prove the lower bound. Next, we introduce terminology related
to {\sc cnf}s. Finally, we define the notion of treewidth.

In this paper by a \emph{set of literals} we mean one that does not
contain an occurrence of a variable and its negation.
For a set $S$ of literals we denote by $Var(S)$ the set of variables
whose literals occur in $S$. If $F$ is a Boolean function
or its representation by a {\sc cnf} or {\sc obdd}, we denote by $Var(F)$
the set of variables of $F$. A truth assignment to $Var(F)$ on which $F$
is true is called a \emph{satisfying assignment} of $F$. A set $S$ of literals
represents the truth assignment to $Var(S)$ where variables occurring
positively in $S$ (i.e. whose literals in $S$ are positive) are assigned with $true$
and the variables occurring negatively are assigned with $false$.
We denote by $F_S$ a function whose set of satisfying assignments consists of $S'$
such that $S \cup S'$ is a satisfying assignment of $F$. We call $F_S$ a \emph{subfunction}
of $F$. In other words, a Boolean function $F'$ is a subfunction of a Boolean function
$F$ is $F'$ can be obtained from $F$ by giving a truth assignment to a subset of variables of $F$.

An {\sc obdd} $Z$ representing a Boolean function $F$ is a directed acyclic graph ({\sc dag}) with one root and two leaves
labelled by $true$ and $false$. 
The internal nodes are labelled with variables of $F$. There is a fixed permutation $SV$ of $Var(F)$
(that is, elements of $Var(F)$ are linearly ordered according to $SV$)
so that the vertices along any path from the root to a leaf are labelled with variables according to this order. 
Each internal vertex is associated with $2$ leaving edges labelled with $true$ and $false$.
Each path $P$ from the root of $Z$ is called a \emph{computational path} and is associated with truth assignment to
the variables labelling all the vertices but the last one. In particular, each variable is assigned with the value labelling
the edge of the path that leaves the corresponding vertex. We denote by $A(P)$ the assignment associated with the 
computational path $P$. The set of all $A(P)$ where $P$ is a computational path ending at the $true$ leaf is precisely
the set of satisfying assignments of $F$. 

\begin{figure}[h]
\centering 
\includegraphics[height=5cm]{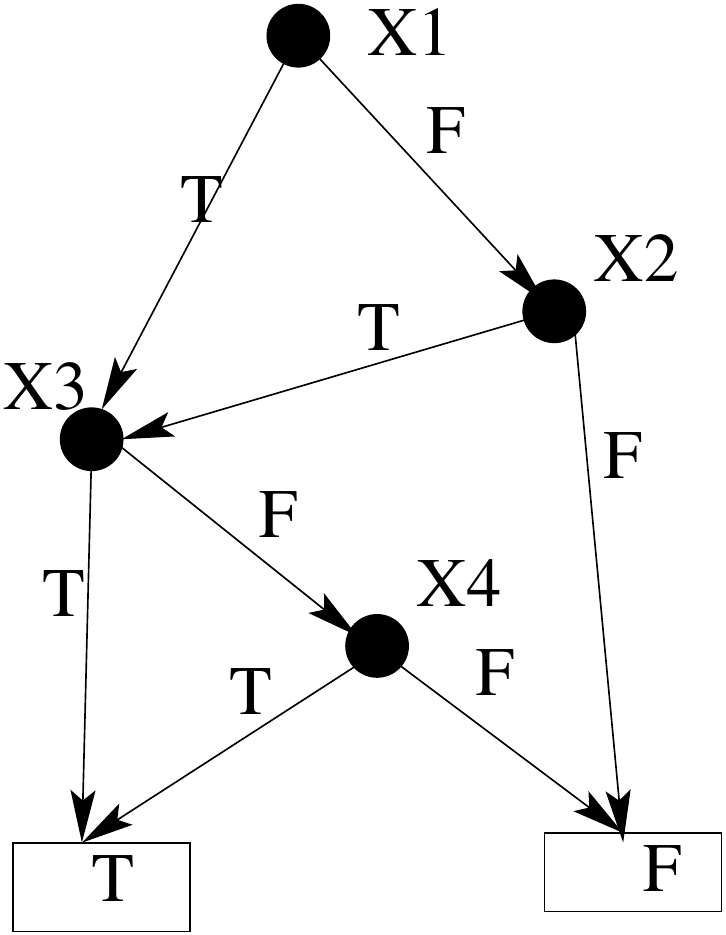}
\caption{An {\sc obdd} for $(x_1 \vee x_2) \wedge(x_3 \vee x_4)$
under permutation $(x_1,x_2,x_3,x_4)$}
\label{OBDDPic}
\end{figure}

Figure \ref{OBDDPic} shows an {\sc obdd} for the function $(x_1 \vee x_2) \wedge (x_3 \vee x_4)$
under the permutation $(x_1,x_2,x_3,x_4)$. Consider the path $P=(x_1,x_2,x_3)$. 
Then $A(P)=\{\neg x_1,x_2\}$. 

In order to obtain the lower bound on the {\sc obdd} size we use a standard approach
of counting subfunctions. See \cite{WegBook} for examples of application of this approach.
This approach is based on the following statement.

\begin{proposition} \label{paths}
Let $F$ be a Boolean function on a set $V$ of variables and let $SV$ be a permutation
of $V$. Partition $SV$ into a prefix $SV_1$
and a suffix $SV_2$ and suppose that the number of distinct subfunctions of $F$ obtained by giving
truth assignments to all the variables of $SV_1$ is at least $x$. Then an {\sc obdd} of $F$ with
the underlying order $SV$ contains at least $x$ nodes. 
\end{proposition}

The standard way to utilize Proposition \ref{paths} is to show that for \emph{any} permutation
$SV$ of $V$ there is a partition of $SV$ into a prefix $SV_1$ and a suffix $SV_2$ such that
the instantiation of variables of $SV_1$ results in at least $x$ different subfunctions.
Then Proposition \ref{paths} immediately implies that $x$ is a lower bound on the size of {\sc obdd}
for \emph{any} underlying order.

Given a {\sc cnf} $F$, its \emph{primal graph} has the set of vertices corresponding to the variables of $F$.
Two vertices are adjacent if and only if there is a clause of $F$ where the
corresponding variables both occur. In the \emph{incidence graph}
of $F$ the vertices are partitioned into those corresponding to the variables of $F$ and those corresponding to its
clauses. A variable vertex is adjacent to a clause vertex if and only if the corresponding variable occurs in the
corresponding clause. 

Given a graph $G$, its \emph{tree decomposition} is a pair $(T,{\bf B})$ where $T$ 
is a tree and ${\bf B}$ is a set of bags $B(t)$ corresponding to the vertices $t$ of $T$.
Each $B(t)$ is a subset of $V(G)$ and the bags obey the rules of \emph{union} (that is, $\bigcup_{t \in V(T)} B(t)=V(G)$),
\emph{containment} (that is, for each $\{u,v\} \in E(G)$ there is $t \in V(t)$ such that $\{u,v\} \subseteq B(t)$),
and \emph{connectedness} (that is for each $u \in V(G)$, the set of all $t$ such that $u \in B(t)$ induces a subtree of $T$).
The \emph{width} of $(T,{\bf B})$ is the size of the largest bag minus one. The treewidth of $G$ is the smallest width of a tree
decomposition of $G$. If $T$ is a path then we use the respective notions of \emph{path decomposition} and \emph{pathwidth}. 

\begin{figure} [h]
\centering
\includegraphics[height=5cm]{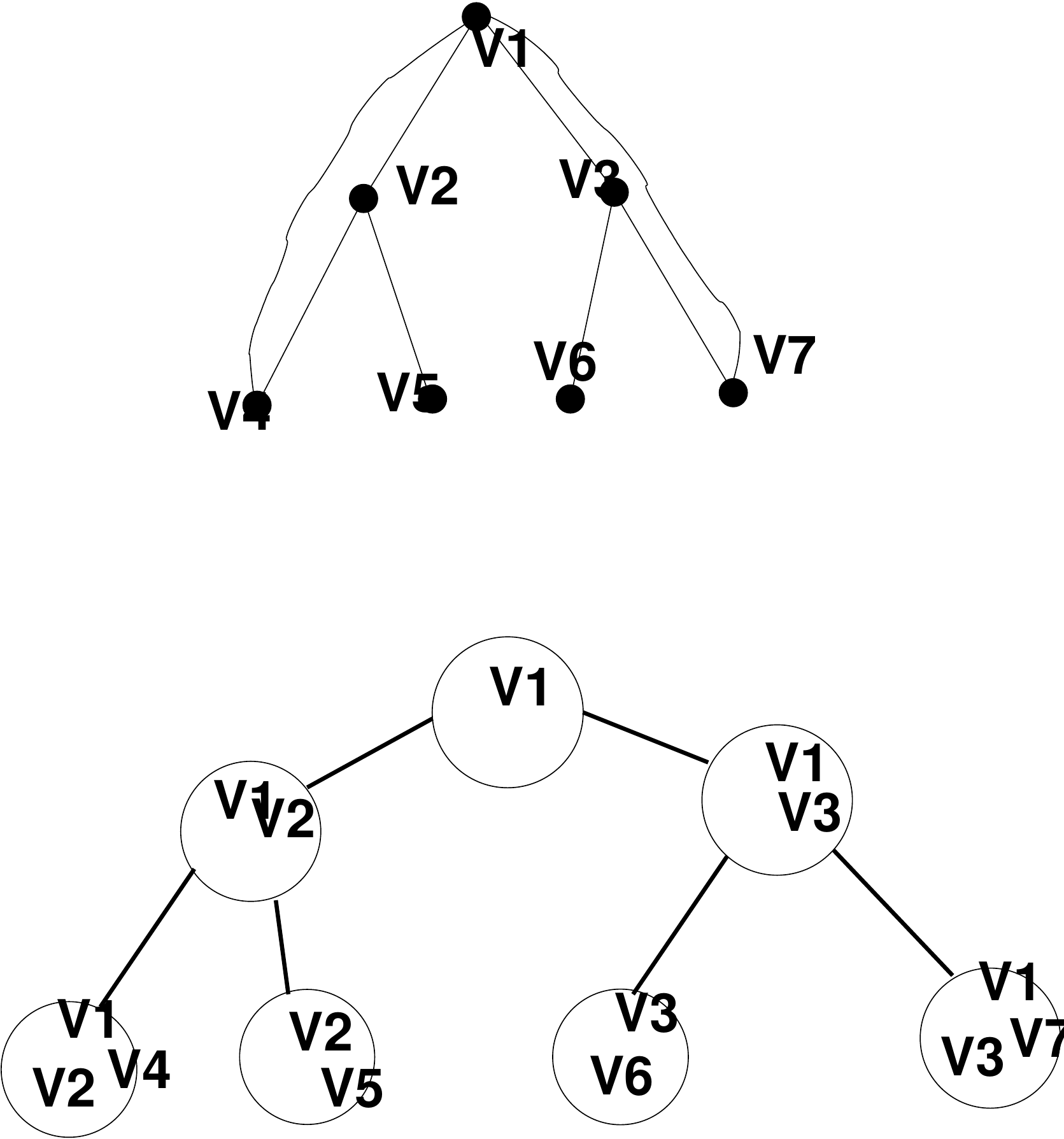}
\caption{A graph and its tree decomposition}
\label{TWDPic}
\end{figure}

Figure \ref{TWDPic} shows a graph and its tree decomposition.
The width of this tree decomposition is $2$ since the size of the largest bag
is $3$.

\section{The lower bound} \label{lbmain}
In this section, given two integers $r$ and $k$ we define a class of {\sc cnf}s, 
roughly speaking, based on complete binary trees of height $r$ where each node is associated 
with a clique of size $k$. Then we prove that the treewidth of the primal graphs of {\sc cnf}s of this
class is linearly bounded by $k$. Further on, we state the main technical theorem (proven in the next
section) that claims that the smallest {\sc obdd} size for {\sc cnf}s of this class exponentially depends
on $rk$. Finally, we re-interpret this lower bound in terms of the number of variables and the treewidth to get
the lower bound announced in the Introduction.

Let $G$ be a graph. A \emph{graph based} {\sc cnf} denoted by $CNF(G)$ is defined as follows. 
The set of variables consists of variables $X_u$ for each $u \in V(G)$
and variables $X_{u,v}=X_{v,u}$ for each $\{u,v\} \in E(G)$. 
The set of clauses consists of clauses $C_{u,v}=C_{v,u}=(X_u \vee X_{u,v} \vee X_v)$
for each $\{u,v\} \in E(G)$. In other words, the variables of $CNF(G)$ correspond to the vertices
and edges of $G$. The clauses correspond to the edges of $G$. 

Denote by $T_r$ a complete binary tree of height $r$.
Let $CT_{r,k}$ be the graph obtained from $T_r$ by associating each vertex
with a clique of size $k$ and, for each edge $\{u,v\}$ of $G$, making all the vertices
of the cliques associated with $u$ and $v$ mutually adjacent.
Denote $CNF(CT_{r,k})$ by $F_{r,k}$. 

\begin{figure}[h]
\centering
\includegraphics[height=5cm]{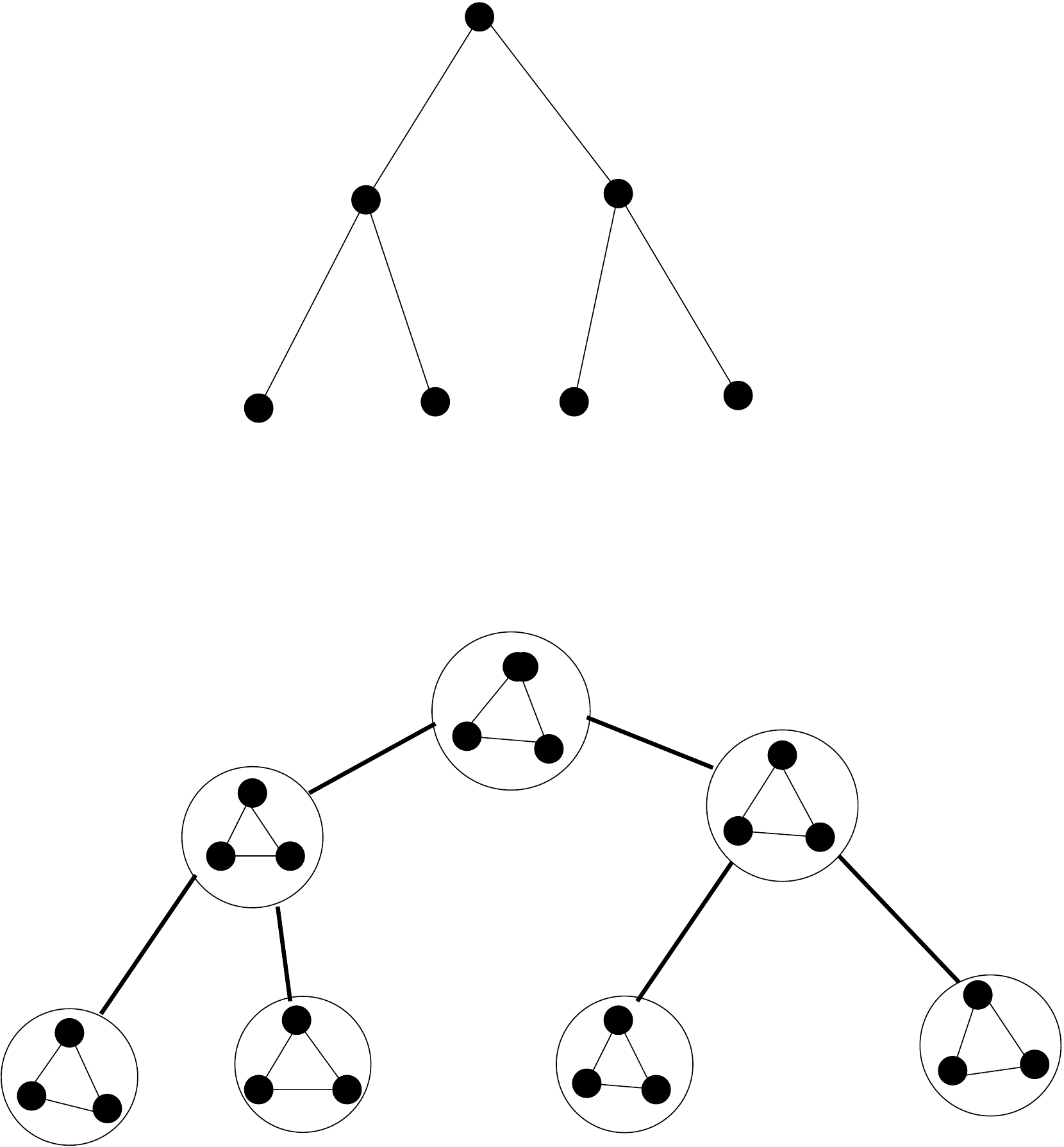}
\caption{$T_2$ and $CT_{2,3}$}
\label{TWPattern}
\end{figure}

Figure \ref{TWPattern} shows $T_2$ and $CT_{2,3}$.
To avoid shading the picture of $CT_{2,3}$ with many edges,
the cliques corresponding to the vertices of $T_2$ are marked
by circles and the bold edges between the circles mean that
that there are edges between all pairs of vertices of the corresponding 
cliques. 

\begin{lemma} \label{twf}
The treewidth of the primal graph of $F_{r,k}$ is at least $k-1$ at most $2k-1$. 
In fact, for $r \geq 1$, this treewidth is exactly $2k-1$. 
\end{lemma}

{\bf Proof.}
The primal graph of $F_{r,k}$ can be obtained from $CT_{r,k}$
by adding one vertex $v_e$ for each edge $e$ of $CT_{r,k}$ and
making this vertex adjacent to the ends of $e$.

The lower bound follows from existence of a clique of size $k$ in $CT_{r,k}$.
Indeed, in any tree decomposition of $CT_{r,k}$, there is a bag containing
all the vertices of such a clique \cite{BodlaenderM93}. Consequently, the
width of any tree decomposition is at least $k-1$. In fact if $r \geq 1$
then $CT_{r,k}$ has a clique of size $2k$ created by cliques of two adjacent
nodes. Hence, due to the same argumentation, the treewidth of $CT_{r,k}$
is at least $2k-1$ for $r \geq 1$. 

For the upper bound,
consider the following tree decomposition $(T,{\bf B})$ of $CT_{r,k}$.
$T$ is just $T_r$. We look upon $T_r$ as a rooted tree, the centre of
$T_r$ being the root. The bag $B(u)$ of each node $u$ contains the clique of
$CT_{r,k}$ corresponding to $u$. In addition, if $u$ is not the root vertex
then $B(u)$ also contains the clique corresponding to the parent of $u$. 
Observe that $(T,{\bf B})$ satisfies the connectivity property.
Indeed, each vertex appears in the bag corresponding to its `own' clique and 
the cliques of its children. Clearly, the set of nodes corresponding to the bags 
induce a connected subgraph. The rest of the tree decomposition properties can
be verified straightforwardly. We conclude that $(T,{\bf B})$ is indeed a tree
decomposition of $CT_{r,k}$.

In order to `upgrade' $(T,{\bf B})$, add ${k \choose 2}$ new adjacent vertices 
to each vertex of $T$. These vertices will correspond to the edges of cliques associated
with the respective nodes of $T_r$. In addition, add $k^2$ new adjacent vertices to each non-root vertex of $T$. 
These vertices will correspond to the edges between the clique associated with the corresponding node of $T_r$
and the clique of its parent.
The bag of each new vertex will contain $v_e$, corresponding to the edge $e$ associated with this bag, 
plus the ends of $e$. A direct inspection shows that this is indeed a tree decomposition
of the primal graph of $F_{r,k}$ and that the size of each bag is at most $2k$.

Notice that for $r \geq 1$ the lower and upper bounds coincide, thus allowing to state
the treewidth precisely. 
$\blacksquare$

The following is the main technical result whose proof is given in the next section. 
\begin{theorem} \label{maintheor}
The size of {\sc obdd} computing $F_{r,k}$ is at least $2^{rk/2}$.
\end{theorem}

The following corollary reformulates the lower bound in terms of the number of variables of $F_{r,k}$ and $k$.

\begin{corollary} \label{reform1}
Let $m$ be the number of variables of $F_{r,k}$.
Then the size of {\sc obdd} computing $F_{r,k}$ is at least
$[g(k)]^{-k/2}m^{k/2}$ where $g(k)=2(k+{k \choose 2}+k^2/4)$
\end{corollary}

{\bf Proof.}
Recall that $T_r$ has $2^{r+1}-1$ nodes.
For each node $a$ of $T_r$, $F_{r,k}$ has $k$ variables corresponding to the
vertices of the clique of $a$ plus ${k \choose 2}$ variables corresponding to the edges
of this clique. In addition, if $a$ is a non-root node then it is associated with $k^2/4$
variables connecting the clique of $a$ with the clique of its parent. Thus each node of
$T_r$ is associated with at most $k+{k \choose 2}+k^2/4$ variables and hence the total number of
variables $m \leq (2^{r+1}-1)*(k+{k \choose 2}+k^2/4) \leq 2^{r+1}*(k+{k \choose 2}+k^2/4)=2^r*g(k)$.
Thus $2^r \geq m/g(k)$. According to Theorem \ref{maintheor}, the
size of an {\sc obdd} computing $F_{r,k}$ is at least
$(2^r)^{k/2} \leq (m/g(k))^{k/2}$ as required. $\blacksquare$

Now we are ready to state the parameterized lower bound for {\sc obdd}s. 

\begin{corollary} \label{paramlower}
There is a function $f$ such that for each $p \geq 1$
there is an infinite sequence of {\sc cnf}s $F_1, F_2 \dots,$ of treewidth at most $p$
of their primal graphs 
such that for each $F_i$ the size of {\sc obdd} 
computing it is at least $f(p)*m^{p/4}$ where $m$ is the number of variables of $F_i$. 
Put it differently, for each fixed $p$, there is a class of {\sc cnf}s
of treewidth at most $p$ of the primal graph for which the {\sc obdd} size is $\Omega(m^{p/4})$. 
\end{corollary}

{\bf Proof.}
For an odd $p$, consider the {\sc cnf}s $F_{r,(p+1)/2}$ for all $r \geq 1$
and for an even $p$, consider the {\sc cnf}s $F_{r,p/2}$ for all $r \geq 1$.
By Lemma \ref{twf}, the treewidth of the primal graph of $F_{r,(p+1)/2}$ is at 
most $p$ and of $F_{r,p/2}$ at most $p-1$. Thus the treewidth requirement is
satisfied regarding these classes. 

By Corollary \ref{reform1}, the {\sc obdd} size is lower-bounded by
$[g((p+1)/2)]^{-(p+1)/2}*m^{(p+1)/4}$ for the former class
and by $[g(p/2)]^{-p/2}*m^{p/4}$ for the latter class.
Observe that $[g((p+1)/2)]^{-(p+1)/2}*m^{p/4}$ is a lower bound for both
these lower bound. Hence, the corollary follows by assuming 
$f(p)=[g((p+1)/2)]^{-(p+1)/2}$.

$\blacksquare$ 

Corollary \ref{paramlower} establishes \emph{parameterized} separation between
{\sc obdd} and {\sc sdd}. The next corollary shows that essentially the same method
can be used to separate {\sc obdd} and {\sc sdd} in the \emph{classical sense}.

\begin{corollary}
There is an infinite family of functions for which the smallest OBDDs
are of size $n^{\Omega(\log n)}$ while there are SDDs of size $O(n^3)$.
\end{corollary}

{\bf Proof}
Consider functions $F_{r,r}$.
Let us compute the number $n$ of variables of $F_{r,r}$. 
Following the calculation as in Corollary \ref{reform1},
we observe that
\begin{equation}
n=(2^{r+1}-1)(\frac{r*(r-1)}{2}+r)+(2^{r+1}-2)\frac{r^2}{4}=
  2^{r}(\frac{3r^2+2r}{2})-\frac{2r^2+r}{2}
\end{equation}

Denote $\frac{3r^2+2r}{2}$ by $p_1$ and $\frac{2r^2+r}{2}$ by $p_2$.
Then 
\begin{equation} \label{maineq}
r=\log \frac{n+p_2}{p_1}
\end{equation}. 

It follows from \eqref{maineq} that for a sufficiently large $r$,
$r \geq \log n-\log p_1 \geq \log n-r/2$ and hence $r \geq (2\log n)/3$.
Then it follows from Theorem \ref{maintheor} that for a sufficiently
large $r$, an {\sc obdd} for $F_{r,r}$ is of size at least 
$2^{4\log^2(n)/(9*2)}=n^{2\log(n)/9}$.

On the other hand, it follows from \eqref{maineq} 
that for a sufficiently large $r$, $r \leq \log(n+p_2) \leq \log(2n)=\log n+1$.
Thus, according to \cite{SDD}, the size of {\sc sdd} for $F_{r,r}$ is bounded by
$O(2^{2log n}n)=O(n^3)$, confirming the required separation.
$\blacksquare$

\section{Proof of Theorem \ref{maintheor}} \label{lbaux1}
The plan of the proof is the following. We introduce the notion of
matching width of a graph. Then we provide two statements regarding this
notion. The first statement (Lemma \ref{cltreemt}) claims a linear in $rk$ lower bound 
for the matching width of graphs $CT_{r,k}$ underlying the considered class
$F_{r,k}$ (the proof of the lemma is provided in the next section).
The second statement (Lemma \ref{manyass}) claims that if a graph $G$ has a matching width $t$
then any permutation of the variables of $CNF(G)$ can be partitioned into 
a suffix and a prefix so that there are at least $2^t$ subfunctions of $CNF(G)$
resulting from instantiation of variables of the prefix. The proof of Lemma \ref{manyass}
constitutes the essential part of this section. 
Finally, we provide a proof of Theorem \ref{maintheor}.
In this proof we notice that according to the approach 
outlined in the Preliminaries section, Lemma \ref{manyass} together
with Proposition \ref{paths} implies that the size of an {\sc obdd} of $CNF(G)$
is at least $2^t$. Taking $CT_{r,k}$ as $G$ and substituting the lower
bound claimed by Lemma \ref{cltreemt}, we obtain the desired lower bound for
$F_{r,k}=CNF(CT_{r,k})$. 

The \emph{matching width} is defined as follows.
Let $SV$ be a \emph{permutation} of the set $V=V(G)$ of vertices of a graph
$G$. 
Let $S_1$ be a \emph{prefix} of $SV$ (i.e. all vertices of $SV \setminus S_1$ are ordered after
$S_1$). Let us call the matching width of $S_1$,
the largest \emph{matching} (that is, a set of edges not having common ends)
consisting of the edges between $S_1$ and $V \setminus S_1$
(we take the liberty to use sequences as sets, the correct use will be always clear
from the context). Further on, the matching width of $SV$ is the largest matching
width of a prefix of $SV$. Finally the matching width of $G$, denoted by $mw(G)$, is the smallest 
matching width of a permutation of $V(G)$. 

\begin{example}
Consider a path of $10$ vertices $v_1, \dots, v_{10}$ so that $v_i$ is adjacent to $v_{i+1}$ for
$1 \leq i<10$. The matching width of permutation $(v_1, \dots, v_{10})$ is $1$ since between any suffix
and prefix there is only one edge. However, the matching width of the permutation 
$(v_1,v_3,v_5,v_7,v_9,v_2,v_4,v_6,v_8,v_{10})$ is $5$ as witnessed by the partition
$\{v_1,v_3,v_5,v_7,v_9\}$ and $\{v_2,v_4,v_6,v_8,v_{10}\}$. Since the matching width of a graph is determined
by the permutation having the smallest matching width, and, since the graph has edges, there cannot be
a permutation of matching width $0$, we conclude that the matching width of this graph is $1$.
\end{example}

\begin{lemma} \label{cltreemt}
For any $r$,
the matching width of $CT_{r,k}$ is at least $rk/2$.
\end{lemma}

The proof of Lemma \ref{cltreemt} is provided is the next section.

{\bf Remark.} The above definition of matching width is a special 
case of a more general notion of \emph{maximum matching width} as defined
in \cite{VaThesis}. In particular our notion of matching width can be seen
as a variant of maximum matching width of \cite{VaThesis} 
where the tree $T$ involved in the definition is a caterpillar. 

We are now showing that for {\sc cnf}s of form $CNF(G)$, a large matching width
of $G$ is sufficient for establishing a strong lower bound. 

\begin{lemma} \label{manyass}
Let $G$ be a graph having matching width $t$.
Denote $CNF(G)$ by $F$. Then any permutation $SF$
of $Var(F)$ has a prefix $SF_1$ such that 
there are at least $2^t$ different functions of form $F_{S_1}$
such that $S_1$ is a truth assignment to the variables of $SF_1$.
\end{lemma}

{\bf Proof.}
Let us partition $Var(F)$ into sets $VV$ of variables corresponding to
the vertices of $G$ and $EV$ of variables corresponding to the edges of 
$G$. Let $SV$ be the permutation of $VV$ ordered in the way as they are ordered
in $SF$. Let $SV_1$ be a prefix of $SV$ \emph{witnessing} the matching width $t$ of $SV$.
(Recall that the matching width of $SV$ is at least the matching width of $G$.)
The word `witnessing' in this context means that there is a matching
$M=\{\{u_1,v_1\}, \dots, \{u_t,v_t\}\}$ between $SV_1$ and $V(G) \setminus SV_1$. 
Let $SF_1$ be the prefix of $SF$ ending with the last element of $SV_1$.
Thus the variables $X_{u_1}, \dots X_{u_t}$ corresponding to $u_1, \dots, u_t$ belong to
$SF_1$ while the variables $X_{v_1}, \dots, X_{v_t}$ corresponding to $v_1, \dots, v_t$
do not. We denote the set of clauses $(X_{u_i} \vee X_{u_i,v_i} \vee X_{v_i})$ by $TCL$.

In the rest of the proof we essentially show that $2^t$ different assignments to variables
$X_{u_1}, \dots X_{u_t}$ produce $2^t$ different subfunctions of $F$ thus confirming the
lemma. Roughly speaking, this is done by showing that by a careful fixing the
assignments to \emph{the rest} of the variables of $SF_1$ we can achieve the
effect that an assignment to $X_{u_i}$ does not `influence' an assignment to $X_{v_j}$ for
$i \neq j$. As a result no two assignments to $X_{u_1}, \dots, X_{u_t}$ can have the same
effect on $X_{v_1}, \dots, X_{v_t}$ and this guarantees that desired large set of subfunctions.

We start from defining a set of $2^t$ assignments for which we then claim that any two assignments
induce two distinct subfunctions of $F$.
In particular, let ${\bf S}$ be the set of all assignments
to the variables of $SF_1$ that assign the variables $X_{u_i,v_i}$  (of course,
those of them that belong to $SF_1$) with $false$
and the rest of variables except $X_{u_1}, \dots, X_{u_t}$ with $true$. 
It is easy to see by construction that ${\bf S}$ is in a natural one-to-one correspondence
with the set of possible assignments to $X_{u_1}, \dots, X_{u_t}$. In particular,
each $S \in {\bf S}$ corresponds to the assignment $A$ to $X_{u_1}, \dots,X_{u_t}$
contained in it. Indeed, the assignments of the rest of the variables are fixed
in ${\bf S}$ by construction. It follows that the size of ${\bf S}$ is $2^t$.

We are going to show that for any distinct $S_1,S_2 \in {\bf S}$, $F_{S_1} \neq F_{S_2}$,
confirming the lemma. Due to the correspondence established above, we can specify $u_i$ such that
$S_1$ and $S_2$ assign $X_{u_i}$ with distinct values. Assume w.l.o.g. that 
$X_{u_i}$ is assigned with $true$ by $S_1$ and with $false$ by $S_2$. 
Observe that $F$ does not have a satisfying assignment including $S_2$ and 
assigning both $X_{u_i,v_i}$ and $X_{v_i}$ with $false$. Indeed, as a result, 
the clause $(X_{u_i} \vee X_{u_i,v_i} \vee X_{v_i})$ is falsified. 
We are going to show that both $X_{u_i,v_i}$ and $X_{v_i}$ can be assigned 
with $false$ in a satisfying assignment of $F$ including $S_1$.
Indeed, assign all the variables of $Var(F) \setminus (Var(S_1) \cup \{X_{u_i,v_i},X_{v_i}\})$ with $true$ and see that the resulting 
assignment together with $S_1$ satisfies all the clauses of $F$.
Indeed, if a clause $(X_u \vee X_{u,v} \vee X_v)$
does not belong to $TCL$ then $X_{u,v}$ is assigned with $true$ 
(by construction, the only `edge' variables assigned by $false$ are $X_{u_i,v_i}$, that is 
those that occur in the clauses of $TCL$) . Furthermore, for any clause
$(X_{u_j} \vee X_{u_j,v_j} \vee X_{v_j})$ of $TCL$ such that $i \neq j$, $X_{v_j}$ is assigned with
$true$. Finally $X_{u_i}$ is assigned with $true$ by $S_1$. It follows that indeed all the clauses of $F$ are satisfied. 

Assume that $X_{u_i,v_i} \notin Var(S_1)$. Then, by the reasoning as above, $F_{S_1}$ has a satisfying assignment 
including $\{\neg X_{u_i,v_i}, \neg X_{v_i}\}$ while $F_{S_2}$ does not implying that $F_{S_1} \neq F_{S_2}$.
Otherwise, if $X_{u_i,v_i} \in Var(S_1)$, it is assigned with $false$ in both $S_1$ and $S_2$, by construction.
It follows that $F_{S_1}$ has a satisfying assignment including $\neg X_{v_i}$ while $F_{S_2}$ does not. It follows 
again that $F_{S_1} \neq F_{S_2}$.
$\blacksquare$

{\bf Remark.} Notice the role of variables $X_{u,v}$ in the proof of Lemma \ref{manyass}.
They allow the values of $X_{u_i}$ to \emph{not influence} the values of $X_{v_j}$ for $i \neq j$
and thus keep the number of different subfunctions up to the desired bound. Due to the same reason, it
is important that the edges $\{u_1,v_1\}, \dots, \{u_r,v_r\}$ constitute a \emph{matching}, i.e. have disjoint
ends. 

{\bf Proof of Theorem \ref{maintheor}}
Lemma \ref{manyass} combined with Proposition \ref{paths}
says that if $G$ has matching width at least $t$ then 
for any permutation of $Var(CNF(G))$ the corresponding {\sc obdd}
has at least $2^t$ nodes. In other words, $2^t$ is a lower bound
on the {\sc obdd} size for $CNF(G)$. Taking $G=CT_{r,k}$
and hence $CNF(G)=F_{r,k}$ and substituting $rk/2$ for $t$ according 
to Lemma \ref{cltreemt}, we obtain a lower bound of $2^{rk/2}$ on the
{\sc obdd} size of $F_{r,k}$, as required. $\blacksquare$ 

\section{Proof of Lemma \ref{cltreemt}} \label{lbaux2}
This section is organized as follows. First, we introduce the 
notion of \emph{induced permutation}. Then we provide proof
of Lemma \ref{cltreemt} for $k=1$. After that, we outline how
to upgrade this special case to a complete proof. Finally, we provide
the complete proof. Note that the proof of the special case 
and the following outline are \emph{technically} redundant. 
However, the reader may find them useful as they provide a 
\emph{sketch} reflecting the proof idea. 

The notion of \emph{induced permutation} is defined as follows.
Let $P_1$ be a permutation of elements of a set $S_1$
and let $S_2 \subseteq S_1$. Then $P_1$ induces a permutation $P_2$ of $S_2$
where the elements of $S_2$ are ordered exactly as they are ordered in $P_1$.
For example, let $S_1=\{1, \dots, 10\}$ and let $S_2$ be the subset of even numbers
of $S_1$. Let $P_1=(1,8,2,9,5,6,7,3,4,10)$. Then $P_2=(8,2,6,4,10)$.

{\bf Proof of the special case of Lemma \ref{cltreemt} for $k=1$}
We are going to prove that for an odd $r$, the matching width of $T_r$ is at least
$(r+1)/2$. For an even $r$ we can simply take a subgraph of $T_r$
isomorphic to $T_{r-1}$ (it is not hard to see that the matching width of a graph is not 
less than the matching width of its subgraph).

The proof goes by induction on $r$. For $r=1$, this is clear,
so consider the case $r>1$. Imagine $T_r$ rooted in the natural way, the root being its centre.
Then $T_r$ has $4$ grandchildren, the subtree rooted by each of them being
$T_{r-2}$. Denote these grandchildren by $T^1, \dots, T^4$.
Let $PV$ be any permutation of the vertices of $T_r$. This permutation induces
respective permutations $PV_1, \dots, PV_4$ of vertices of $T^1, \dots, T^4$
being ordered exactly as in $PV$. By the induction assumption, we know that
each of $PV_1, \dots, PV_4$ can be partitioned into a prefix and a suffix
so that the edges between the prefix and the suffix induce graph having
matching of size at least $(r-1)/2$. Each of these prefixes naturally corresponds
to the prefix of $PV$ ending with the same vertex. Since $PV_1, \dots, PV_4$ are 
pairwise disjoint, this correspondence supplies $4$ \emph{distinct} prefixes 
$P^*_1, \dots, PV^*_4$ of $PV$.
Moreover, for each $PV^*_i$ we know that the graph $G^*_i$ induced by the edges between the
vertices of $PV^*_i$ and the rest of the vertices has a matching of size $(r-1)/2$ consisting 
\emph{only} of the edges of $T^i$. In order to `upgrade' this
matching by $1$ and hence to reach the required size of $(r+1)/2$, all we need to show is that
in an least one $G^*_i$ there is an edge both ends are not vertices of $T^i$ and hence this
edge can be safely added to the matching. 

At this point we make a notational assumption that does not lead to loss of generality 
and is convenient for the further exposition. By construction, $PV^*_1, \dots, PV^*_4$
are linearly ordered by containment and we assume w.l.o.g. that the ordering is by the increasing 
order of the subscript, that is $PV^*_1 \subset PV^*_2 \subset PV^*_3 \subset PV^*_4$. 
We claim that the upgrade to the matching as specified above is possible for $PV^*_2$.

Indeed, observe that $T_r \setminus T^2$ is a connected graph. Thus all we need to show
is that at least one vertex of $T_r \setminus T^2$ gets into $PV^*_2$ and at least one
vertex of $T_r \setminus T^2$ gets outside $PV^*_2$, that is in $V(T_r) \setminus PV^*_2$.

For the former, recall that $PV^*_1 \subset PV^*_2$ and that by construction, $PV^*_1$
contains $(r-1)/2$ vertices of $T^1$ being a subgraph of $T_r \setminus T^2$.
Thus we conclude that $PV^*_2$ contains vertices of $T_r \setminus T^2$
For the latter, observe that since $PV^*_2 \subset PV^*_3$, $V(T_r) \setminus PV^*_3 \subset 
V(T_r) \setminus PV^*_2$. Furthermore, by construction, $V(T_r) \setminus PV^*_3$ contains
$(r-1)/2$ vertices of $T^3$ being a subgraph of $T_r \setminus T^2$. 
Thus we conclude that $V(T_r) \setminus PV^*_2$ contains vertices of $T_r \setminus T^2$ as well,
thus finishing the proof. $\blacksquare$

A proof for the general case of Lemma \ref{cltreemt} proceeds by induction on $r$ similarly
to the special case above. Of course we need to keep in mind that instead of nodes of $T_r$
we have cliques of size $k$. The consequence of this substitution is that at the inductive step
of moving from $T_{r-2}$ to $T_r$ we can increase the matching width by $k$ rather than by $1$
as above. The auxiliary Lemma \ref{kmatching} allows us to demonstrate the possibility of this
upgrade essentially in the same way as we did for $k=1$: we just show that the considered prefix 
and suffix of the given permutation both contain at least $k$ vertices outside the grandchild
serving the part of the matching guaranteed by the induction assumption. 

\begin{lemma} \label{kmatching}
Let $T$ be a tree with at least $2$ nodes and let $k$ be a positive integer.
Let $CT$ be a graph obtained from $T$ by associating
with each vertex of $T$ a clique of an arbitrary size $k' \geq k$
and making the vertices of cliques associated with adjacent vertices
of $T$ mutually adjacent. Let $W,B$ standing for 'white'
and 'black' be a partition of $V(CT)$ such that $|W| \geq k$
and $|B| \geq k$. Then $CT$ has a matching of size $k$ formed by edges
with one white and one black end.
\end{lemma}

{\bf Proof.}
The proof is by induction
on the number of nodes of $T$. It is clearly true when there are $2$ nodes.
Assume that the tree has $n>2$ nodes and let $u$ be a leaf of $T$
and $v$ be its only neighbour. 

Let $k' \geq k$ be the size of the clique $VU$ associated with $u$ in $CT$.
Assume w.l.o.g. that $|W \cap VU| \leq |B \cap VU|$. Denote $|W \cap VU|$
by $k_1$. Clearly, the $k_1$ vertices of $W \cap VU$ can be matched with the
vertices of $B \cap VU$. If $k_1 \geq k$, we are done. Next, 
if $|B \setminus VU| \geq k-k_1$, then the lemma follows by induction assumption
applied on $T \setminus u$.

Consider the remaining possibility where $|B \setminus VU|=k-k_1-t$ for some
$t>0$. Observe that $t \leq k'-2k_1$. Indeed, the total number of vertices of 
$B$ is $k'-k_1+k-k_1-t$ so, $t>k'-2k_1$ will imply $|B|<k$, a contradiction.

Let $VV$ be the clique associated 
with the neighbour $v$ of $u$. It follows from our assumption that $|W \cap VV| \geq k_1+t$
because at most $k-k_1-t$ vertices of $VV$ can be black. Match $k_1$ vertices
of $W \cap VU$ with vertices of $B \cap VU$ (this is possible due to our assumption that
$|W \cap VU| \leq |B \cap VU|$). Match $t$ unmatched vertices of 
$B \cap VU$ (there are $k'-2k_1$ unmatched vertices of $B \cap VU$ and we have just shown
that $t \leq k'-2k_1$) with $t$ vertices of $W \cap VV$. We are in the situation where
in $G \setminus u$ there are at least $k-k_1-t$ vertices of $W$, at least
$k-k_1-t$ vertices of $B$ and the size of each associated clique is clearly at 
least $k-k_1-t$. Hence, the lemma follows by the induction assumption.
$\blacksquare$

{\bf Proof of Lemma \ref{cltreemt}.}
We prove that for an odd $r$, the matching width of $CT_{r,k}$ is at least
$(r+1)k/2$. For an even $r$, it will be enough to consider a subgraph of
$CT_{r,k}$ being isomorphic to $CT_{r-1,k}$.
The proof is by induction on $r$.
Assume first that $r=1$. Then the lemma holds according to
Lemma \ref{kmatching}.

For $r>1$, let us view $T_r$ as a rooted tree with its centre $rt$ being the 
root. Let $T^1, \dots, T^4$ be the $4$ subtrees of $T_r$ rooted by the `grandchildren'
of $rt$. Let $K_1, \dots, K_4$ be the subgraphs of $CT_{r,k}$ `corresponding' to
$T^1, \dots, T^4$. That is, each $K_i$ is a subgraph of $CT_{r,k}$ 
induced by (the vertices of) cliques associated with the vertices
of $T^i$. It is not hard to see that each $T^i$ is isomorphic to $T_{r-2}$ and each
$K_i$ is isomorphic to $CT_{r-2,k}$ and that $K_1, \dots, K_4$ are pairwise disjoint. 

Let $PV$ be an arbitrary permutation of $V(CT_{r,k})$.
Let $PV_1, \dots, PV_4$ be the respective permutations of $V(K_1), \dots, V(K_4)$
induced by $PV$. By the induction assumption for each $PV_i$ there is a prefix
$PV'_i$ such that the edges of $K_i$ with one end in $PV'_i$ and the other end in
$PV_i \setminus PV'_i$ induce a graph having matching of size at least $(r-1)k/2$.
Let $u_1, \dots, u_4$ be the last vertices of $PV'_1, \dots PV'_4$, respectively.
Assume w.l.o.g. that these vertices occur in $PV$ in exactly this order.
Let $PV'$ be the prefix of $PV$ with final vertex $u_2$. We are going to show that
the subgraph of $CT_{r,k}$ induced by the edges between $PV'$ and $PV \setminus PV'$
has matching of size at least $(r+1)k/2$. In fact, as specified above, we already have 
matching of size $(r-1)k/2$ if we confine ourself to the edges between $PV' \cap PV_2$
and $(PV \setminus PV') \cap PV_2$. Thus, it only remains to show the existence of matching
of size $k$ in the subgraph of $CT_{r,k}$ induced by the edges between
$PV^*_1=PV' \setminus PV_2$ and $PV^*_2=(PV \setminus PV') \setminus PV_2$. 
Observe that $PV^*_1,PV^*_2$ is a partition of vertices of $CT_{r,k} \setminus K_2$.
Therefore, it is sufficient to show that $|PV^*_1| \geq k$ and $|PV^*_2| \geq k$ and then
the existence of the desired matching of size $k$ will follow from Lemma \ref{kmatching}.

Due to our assumption that $u_1$ precedes $u_2$ in $PV$, it follows that $PV'_1$ is contained in $PV'$.
Moreover, since $K_1$ and $K_2$ are disjoint, $PV'_1$ is disjoint with $PV_2$ and hence $PV'_1 \subseteq PV^*_1$.
Recall that by the induction assumption, the vertices of $PV'_1$ serve as ends
of a matching of size $(r-1)k/2$ with no two vertices sharing the same edge of the matching.
That is $|PV'_1| \geq (r-1)k/2$. Since $r>1$ by assumption, we conclude that $|PV'_1| \geq k$ and
hence $|PV^*_1| \geq k$.

The proof that $|PV^*_2| \geq k$ is symmetrical. By our assumption, $u_2$ precedes $u_3$ is $PV$ and
hence $PV_3 \setminus PV'_3$ is contained in $PV \setminus PV'$ and due to the disjointness of $K_2$ and
$K_3$, $PV_3 \setminus PV'_3$ is in fact contained in $PV^*_2$. That $|PV_3 \setminus PV'_3| \geq k$
is derived analogously to the proof that $|PV'_1| \geq k$. $\blacksquare$

\section{{\sc obdd}s parameterized by the treewidth of the incidence graph} \label{twinc}
Recall that the incidence graph of the given {\sc cnf} $F$ has the set of vertices
corresponding to its variables and clauses and a variable vertex is adjacent to a clause vertex if and 
only if the corresponding variable occurs in the corresponding clause. 
The upper bound of \cite{VardiTWD} does not straightforwardly apply to the 
case of incidence graphs because there are classes of {\sc cnf}s having constant treewidth of the 
incidence graph and unbounded treewidth of the primal graph. Indeed, consider, for example a {\sc cnf}
with one large clause. Nevertheless, we show in this section that the $O(n^k)$ upper bound on the
size of {\sc obdd} holds if $k$ is the treewidth of the incidence graph of the considered {\sc cnf}.

As in \cite{VardiTWD}, we show that if $p$ is the pathwidth of the
\emph{incidence graph} $G$ of the given {\sc cnf} $F$ then the function of $F$
can be realized by an {\sc obdd} of size $O(2^pn)$ implying (through the $k=O(p*logn)$) the
$O(n^k)$ upper bound where $k$ is the treewidth of $G$. The resulting {\sc obdd} is seen 
as a {\sc dag} whose nodes are partitioned into layers, each layer consisting of nodes labelled
by the same variable. The main technical lemma shows that under the right permutation of variables
the nodes of each layer correspond to $O(2^p)$ subfunctions of $F$. Consequently, $O(2^p)$
nodes per layer are sufficient, which in turn, immediately implies the desired upper bound.

Let us start from fixing the notation. 
Let $F$ be a {\sc cnf} and $G$ be its incidence graph, 
whose nodes are $X_1, \dots, X_n$ (corresponding to the variables of $F$)
and $C_1, \dots, C_m$ (corresponding to the clauses of $F$) and $X_i$ and adjacent
to $C_j$ if and only if $X_i$ occurs in $C_j$ (for the sake of brevity, we identify 
the vertices of $G$ with the corresponding variables and clauses). 
Let $(P,{\bf B})$ be a path decomposition of $G$.  Fix an end
vertex of $P$ and enumerate the vertices of $P$ along the path starting from
this fixed vertex. Let $v_1, \dots, v_r$ be the enumeration.
For each $X_i$, let $f(X_i)$ be the smallest $j$ such that $X_i \in B(v_j)$.
We call a linear ordering $SV$ of $X_1, \dots, X_n$ such $X_i<X_j$ whenever
$f(X_i)<f(X_j)$ an ordering \emph{respecting} $f$. 

Now we are ready to prove the main technical lemma.

\begin{lemma} \label{layersize}
Let $SV$ be an ordering respecting $f$.
Let $SV_1$ be a prefix of $SV$. Then the number of distinct
$F_S$ such that $S$ is an assignment to $SV_1$ is at most
$1+2*2^p$ where $p$ is the width of $(P,{\bf B})$.
\end{lemma} 

\begin{figure}[h] 
\centering
\includegraphics[height=6cm]{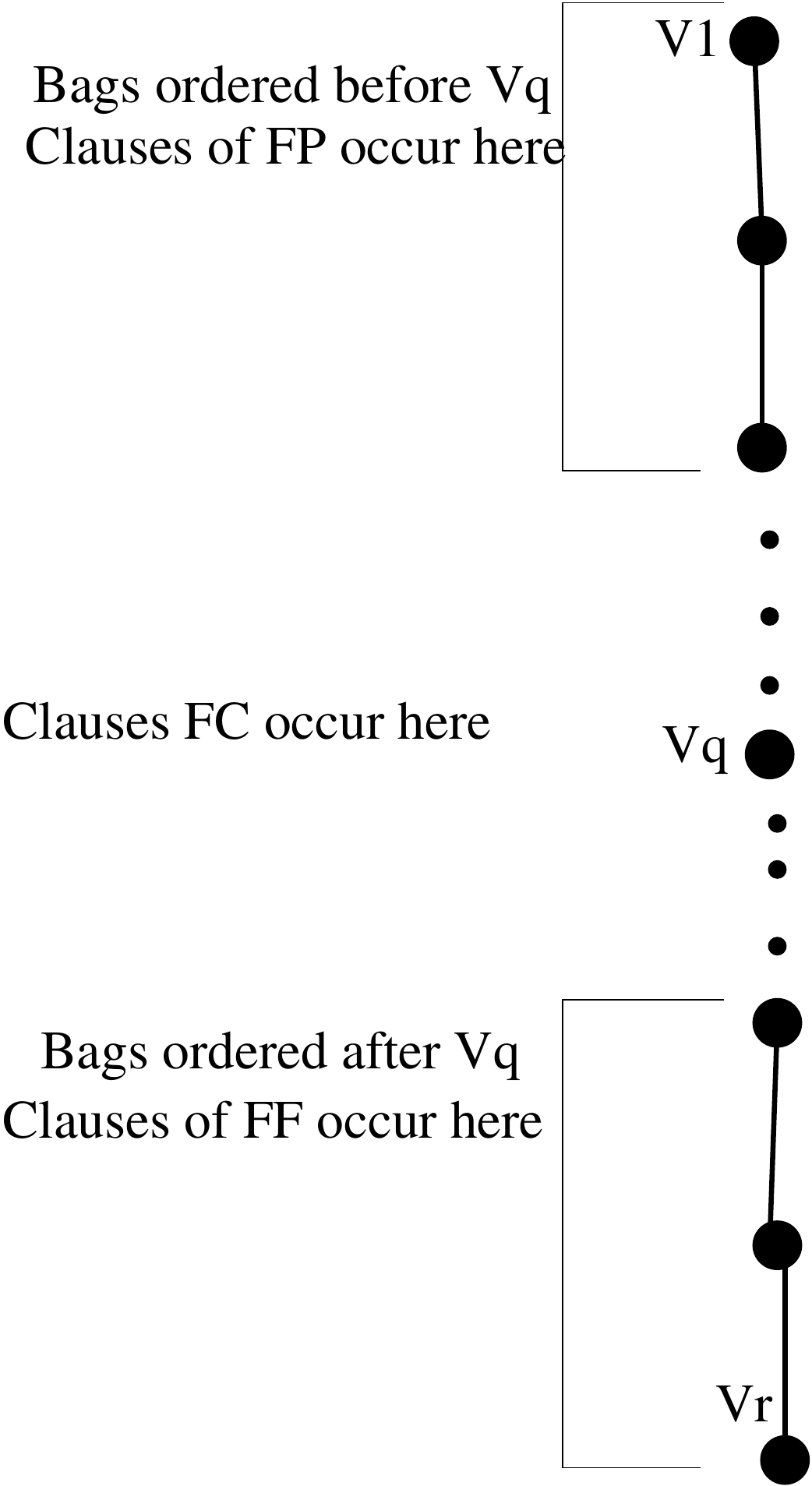}
\caption{Black circles correspond to vertices $v_1, \dots, v_r$ of $P$. Clauses of $FP$
and $FF$ cannot belong to $B(v_q)$ by definition. Suppose that a clause $C \in FP$ belongs 
to $FF$. Then $C$ belongs to a bag of a vertex \emph{above} $v_q$ and to a bag of a vertex \emph{below} $v_q$.
By the connectivity property, $C$ must belong to $B(v_q)$, a contradiction.}
\label{FPFCFFPic}
\end{figure}

{\bf Proof.} 
Let $X$ be the last variable of $SV_1$. Denote $f(X)$ by $q$. 
We assume w.l.o.g. that all the clauses of $F$ are pairwise distinct
and hence identify a {\sc cnf} with its set of clauses. 
Partition $F$ into three sets of clauses: $FP$, consisting of those that 
appear in some $B(v_j)$ for $j<q$ and do not appear in $B(v_q)$; $FC$, 
consisting of those that appear in $B(V_q)$ and $FF$ consisting of those
that appear in $B(v_j)$ for some $j>q$ and do not appear in $B(V_q)$.
Observe that this is indeed a partition of clauses.
Indeed, otherwise $FP \cap FF \neq \emptyset$ as all other possibilities
contradict the definition of the sets $FP,FC,FF$. 
Then due to the connectedness property of $(P,{\bf B})$,
either $FP \cap B(v_q) \neq \emptyset$ or $FF \cap B(v_q) \neq \emptyset$.
However, both these possibilities contradict the definition of $FP$ and $FF$.
We conclude that $FP,FC,FF$ indeed partition the clauses of $F$. For a visual 
justification of their disjointness, see Figure \ref{FPFCFFPic}.

Denote by ${\bf FS}$ the set of all functions $F_S$ such that $S$ is an assignment
to $SV_1$. Denote by ${\bf FPS}$, ${\bf FCS}$, ${\bf FFS}$ the analogous sets regarding
$FP$, $FC$, and $FF$, respectively. 

Let us compute the sizes of the latter $3$ sets. Let $C$ be a clause of $FP$. 
By definition $Var(C)$ is a subset of variables appearing in the bags $B(v_j)$
for $j<q$. By definition, these variables are ordered \emph{before} $X$. It follows
that $Var(C) \subset Var(SV_1)$ and hence any assignment to $SV_1$ either satisfies or
falsifies $C$. Consequently $FP_S$ is either $true$ or $false$.

It is not hard to see that $FC_S$ is obtained from $FC$ by removal of all the clauses
that are satisfied by $S$ and removal of the occurrences of $Var(S)$ from the rest of the 
clauses. It follows that if $FC_{S_1}$ and $FC_{S_2}$ have the same set of satisfied 
clauses then $FC_{S_1}=FC_{S_2}$ in other words, $FC_S$ is completely determined by
a set of satisfied clauses. Hence $|{\bf FCS}|$ is bounded above by the number of 
subsets of clauses of $FCS$, i.e. it is at most $2^{t_1}$ where $t_1$ is the number of 
clauses of $FCS$.

Finally let $SV^*=SV_1 \cap Var(FF)$. It is not hard to see that for an assignment $S$
to $SV_1$, $FF_S$ is completely determined by the subset of $S$ assigning the variables
of $SV^*$. Therefore, the number of distinct functions $FF_S$ is at most
as the number of distinct assignments to $SV^*$, which is $2^{t_2}$ where $t_2=|SV^*|$.

Let $S$ be an assignment on $SV_1$. It is not hard to see that 
$F_S=FP_S \wedge FC_S \wedge FF_S$. If $FP_S=false$ then $F_S=false$. Otherwise,
$FP_S=true$ and hence $F_S=FC_S \wedge FF_S$. In other words, $F_S$ is either false
or there are $F_1 \in {\bf FCS}$ and $F_2 \in {\bf FFS}$ such that $F_S=F_1 \wedge F_2$.
That is $|{\bf FS}| \leq 1+2^{t_1+t_2}$. 

We claim that $t_1+t_2 \leq p+1$ implying the lemma. Indeed, the clauses of $FC$ all belong to
$B(v_q)$ by definition. Observe that $SV^* \subseteq B(v_q)$ as well.
Indeed, let $Y \in SV^*$. Since $Y$ is either $X$ or ordered before $X$,
there must be $j_1 \leq q$ such that $Y \in B(v_{j_1})$. On the other hand,
by definition of $FF$, there must be $j_2>q$ such that $Y \in B(v_{j_2})$.
By the connectedness property $Y \in B(v_q)$. Since $FC$ and $SV^*$ are clearly
disjoint being a set of `clause vertices' and a set of `variable vertices', 
the size of their union is the sum of their sizes and the size of their union cannot be larger
that $|B(v_q)| \leq p+1$, as required. $\blacksquare$

The upper bound can now be formally stated. 

\begin{theorem}
Let $F$ be a {\sc cnf} with $n$ variables and the pathwidth $p$ of its incidence graph. 
Then $F$ can be compiled into an {\sc obdd} of size $O(2^pn)$. 
\end{theorem}

{\bf Proof.}
In fact we prove that the $O(2^pn)$ upper bound holds even for \emph{uniform} {\sc obdd}s
where each path from the root to a leaf includes \emph{all} the variables. 
Notice that the uniformity is not required by the definition of the {\sc obdd}, only
the order of variables along a computational path is essential. For instance, the {\sc obdd}
shown in Figure \ref{OBDDPic} is not uniform. 

Let $SV$ be an ordering respecting $f$ as above. Let $Z$ be a smallest possible uniform {\sc obdd} of $F$ 
with $SV$ being the underlying ordering.  It is well known that the subgraph of $Z$ induced by any internal node $u$
and all the vertices reachable from $u$ (the labels on vertices and edges are retained) is an {\sc obdd}
whose function is $F_{A(P)}$ where $P$ is an arbitrary path from the root to $u$ 
(recall that $A(P)$ denotes the assignment associated with $P$). 
Moreover, the minimality of $Z$ implies that all the nodes marked with the same variable represent distinct functions.
Indeed, if there are $2$ nodes representing the same function then one of them can be removed, with the
in-edges of the removed node becoming the in-edges of another node associated with the same function and with possible
removal of some nodes that become not reachable from the root. This produces another uniform {\sc obdd} implementing
the same function and having a smaller size in contradiction to the minimality of $Z$.

By construction the function of a node labelled with a variable $x$ of $F$ is a subfunction of
$F$ obtained by an assignment to the variables preceding $x$ in $SV$. According to Lemma \ref{layersize}
the number of such subfunctions is $O(2^p)$. Since distinct nodes labelled by $x$ are associated
with distinct subfunctions, there are $O(2^p)$ nodes labelled by $x$. Multiplying this by the number
$n$ of variables of $F$, we obtain the desired $O(2^pn)$ bound on the number of nodes of $Z$. 
$\blacksquare$

\begin{corollary}
A {\sc cnf} with $n$ variables and having treewidth $k$ can be compiled into an {\sc obdd}
of size $O(n^k)$.
\end{corollary}

We close this section with discussion of yet another parameter of {\sc cnf}s, introduced in \cite{HuDar}, 
whose fixed value guarantees a linear size {\sc obdd}. 
In \cite{HuDar} this parameter has not been given a name so, let us name it \emph{combined width}.
Let $SV$ be a linear ordering on variables of
the given {\sc cnf} $F$. For each variable $x$ in this ordering we define the \emph{cutwidth} of $x$ (w.r.t. to $SV$)
as the number of clauses with one variable being either $x$ or ordered before $x$ and one variable ordered after $x$ in $SV$. 
Further on, we define the \emph{pathwidth} of $x$ (w.r.t. to $SV$) as the number of variables that are either $x$ or ordered
before $x$ that occur in clauses having at least one occurrence of a variable ordered after $x$. 
The combined width of $x$ is the minimum of the cutwidth and the pathwdith of $x$. The combined width
of $SV$ is the maximum over all the combined widths of the variables. Finally, the combined width of $F$ 
is the minimum of combined widths of all possible orders of the variables of $F$. It is shown in \cite{HuDar}
that a {\sc cnf} of combined width $w$ can be complied into an {\sc obdd} of size $O(2^wn)$. 

The combined width of $F$ is a mixture of two parameters of the primal graph of $F$: the cutwidth
(maximum cutwidth of a variable in the given permutation taken minimum over all permutations)
and the pathwidth. Moreover, the combined width is not just their minimum but can in fact be 
much smaller than both cutwidth and pathwidth. Consider for example a {\sc cnf}
$F=F_1 \wedge F_2$ where $F_1$ and $F_2$ are {\sc cnf}s defined as follows.
$F_1=(x \vee x_1) \wedge \dots \wedge (x \vee x_{m})$  and $F_2=(y_1, \dots, y_m)$
We assume that the variables of $F_1$ are disjoint with the variables of $F_2$ and that $m$
can be arbitrarily large. The primal graph of $F_1$ has a large cutwidth. Indeed, for any ordering
of variables of $F_1$ there is a subset $V'$ of $\{x_1, \dots, x_m\}$ of size at least $m/2$ that are
either all smaller than $x$ or all larger than $x$. Specify a variable $y \in V'$ that is a 'median' of $V'$
according to the considered order. Then the cutwidth of this variable will be about $m/4$.
Furthermore, the pathwidth of the primal graph of $F_2$ is large because this graph is just one big clique.
On the other hand, the combined width of $F_1$ and $F_2$ is small. Indeed, order the variables as follows:
$x,x_1, \dots, x_m,y_1, \dots, y_m$. Then the pathwidth index of the first $m+1$ variables is $1$ and hence
the combined width will be at most $1$ as well. Further, the cutwidth of the last $m$ variable is $1$ and hence
the combined width of these variables is $1$ as well. Thus the combined width of this order is $1$ and hence
the combined width of $F_1 \wedge F_2$ is at most $1$ which is clearly much smaller than the minimum of the 
pathwdith and the cutwidth of $F$ (determined by the respective connected components of the primal graph 
of $F$). We leave the relationship between the incidence graph treewidth and the combined width as an open question. 


\begin{thebibliography}{10}

\bibitem{BodlaenderM93}
Hans~L. Bodlaender and Rolf~H. M{\"o}hring.
\newblock The pathwidth and treewidth of cographs.
\newblock {\em SIAM J. Discrete Math.}, 6(2):181--188, 1993.

\bibitem{DynSDD}
Arthur Choi and Adnan Darwiche.
\newblock Dynamic minimization of sentential decision diagrams.
\newblock In {\em AAAI}, 2013.

\bibitem{DarwicheJACM}
Adnan Darwiche.
\newblock Decomposable negation normal form.
\newblock {\em J. ACM}, 48(4):608--647, 2001.

\bibitem{SDD}
Adnan Darwiche.
\newblock {S}{D}{D}: A new canonical representation of propositional knowledge
  bases.
\newblock In {\em IJCAI}, pages 819--826, 2011.

\bibitem{DerMar}
Adnan Darwiche and Pierre Marquis.
\newblock A knowledge compilation map.
\newblock {\em J. Artif. Intell. Res. (JAIR)}, 17:229--264, 2002.

\bibitem{VardiTWD}
Andrea Ferrara, Guoqiang Pan, and Moshe~Y. Vardi.
\newblock Treewidth in verification: Local vs. global.
\newblock In {\em LPAR}, pages 489--503, 2005.

\bibitem{HuDar}
Jinbo Huang and Adnan Darwiche.
\newblock Using dpll for efficient obdd construction.
\newblock In {\em SAT}, 2004.

\bibitem{OBDDTWJha}
Abhay~Kumar Jha and Dan Suciu.
\newblock On the tractability of query compilation and bounded treewidth.
\newblock In {\em ICDT}, pages 249--261, 2012.

\bibitem{Yukna}
Stasys Jukna.
\newblock {\em Boolean Function Complexity: Advances and Frontiers}.
\newblock Springer-Verlag, 2012.

\bibitem{McMillan94}
Kenneth~L. McMillan.
\newblock Hierarchical representations of discrete functions, with application
  to model checking.
\newblock In {\em CAV}, pages 41--54, 1994.

\bibitem{SubbaTree}
Sathiamoorthy Subbarayan, Lucas Bordeaux, and Youssef Hamadi.
\newblock Knowledge compilation properties of tree-of-{B}{D}{D}s.
\newblock In {\em AAAI}, pages 502--507, 2007.

\bibitem{VaThesis}
Martin Vatshelle.
\newblock {\em New width parameters of graphs}.
\newblock PhD thesis, Department of Informatics, University of Bergen, 2012.

\bibitem{WegBook}
Ingo Wegener.
\newblock {\em Branching Programs and Binary Decision Diagrams}.
\newblock SIAM, 2000.

\bibitem{WegSurvey}
Ingo Wegener.
\newblock Bdds--design, analysis, complexity, and applications.
\newblock {\em Discrete Applied Mathematics}, 138(1-2):229--251, 2004.

\bibitem{SDDvsOBDD}
Yexiang Xue, Arthur Choi, and Adnan Darwiche.
\newblock Basing decisions on sentences in decision diagrams.
\newblock In {\em AAAI}, 2012.

\end{thebibliography}

\end{document}